\documentclass[acmsmall]{acmart}
\usepackage{kotex}
\usepackage{multirow}
\usepackage{color, colortbl, xcolor}

\definecolor{LightGray}{gray}{0.97}
\usepackage{tcolorbox}
\usepackage{longtable}
\usepackage{capt-of}
\usepackage{balance,bm}
\usepackage{color, colortbl, xcolor}
\definecolor{LightGray}{gray}{0.97}

\usepackage{url}
\usepackage{hyperref}
\hypersetup{
    colorlinks=true,
}

\usepackage{subcaption}
\usepackage{booktabs} 
\usepackage{graphicx}
\usepackage{textcomp}
\usepackage{color,soul}
\usepackage{bm}
\usepackage{multirow}
\usepackage{wrapfig}
\usepackage{balance}
\usepackage{graphicx}  
\usepackage{enumitem}

\usepackage{colortbl}
\usepackage{arydshln}
\usepackage [english]{babel}
\usepackage [autostyle, english = american]{csquotes}
\definecolor{linkColor}{RGB}{6,125,233}
\definecolor{green}{rgb}{0.0, 0.65, 0.31}
\definecolor{bleudefrance}{rgb}{0.19, 0.55, 0.91}
\definecolor{ceruleanblue}{rgb}{0.16, 0.32, 0.75}
\definecolor{grey}{HTML}{969696}
\definecolor{violet}{HTML}{756bb1}
\definecolor{dgrey}{HTML}{01665e}
\definecolor{lgrey}{HTML}{5ab4ac}
\definecolor{dgreen}{HTML}{005a32}
\definecolor{purple}{HTML}{ae017e}

 \definecolor{editCol}{HTML}{000000}
\definecolor{maskCol}{HTML}{c51b7d}
\definecolor{lrColor}{HTML}{8856a7}
\definecolor{trColor}{HTML}{d01c8b}
\definecolor{ctColor}{HTML}{4dac26}
\definecolor{brickred}{HTML}{f03b20}
\definecolor{improveCol}{HTML}{253494}
\definecolor{worsenCol}{HTML}{d7191c}
\definecolor{DarkBlue}{HTML}{00008B}
\definecolor{mscolor}{HTML}{01665e}
\definecolor{nmscolor}{HTML}{bf812d}
\definecolor{lgreen}{HTML}{ccece6}
\definecolor{dolive}{HTML}{308014}

\definecolor{maskCol}{HTML}{c51b7d}
\definecolor{lrColor}{HTML}{8856a7}
\definecolor{trColor}{HTML}{d01c8b}
\definecolor{ctColor}{HTML}{4dac26}
\definecolor{brickred}{HTML}{f03b20}
\definecolor{improveCol}{HTML}{253494}
\definecolor{worsenCol}{HTML}{d7191c}
\definecolor{lgreen}{HTML}{e0f3db}
\definecolor{dpink}{HTML}{CD1076}
\definecolor{pink}{HTML}{FED2D2}
\definecolor{soothinggreen}{HTML}{4dac26}
\definecolor{darkred}{HTML}{8B0000}

\definecolor{dblue}{HTML}{104E8B}
\definecolor{violet}{HTML}{8A2BE2}
\definecolor{mscolor}{HTML}{01665e}
\definecolor{nmscolor}{HTML}{d8b365}
\definecolor{deepgrey}{HTML}{525252}
\definecolor{dslate}{HTML}{2F4F4F}
\definecolor{dolive}{HTML}{556B2F}
\definecolor{teal}{HTML}{388E8E}
\definecolor{mscolor}{HTML}{01665e}
\definecolor{nmscolor}{HTML}{d8b365}

\definecolor{aicolor}{HTML}{018571}
\definecolor{occolor}{HTML}{ff7799}

\definecolor{srcolor}{HTML}{e34a33}
\definecolor{smcolor}{HTML}{253494}
\definecolor{srsmcolor}{HTML}{7fcdbb}
\definecolor{bothcolor}{HTML}{fe9929}
\definecolor{onecolor}{HTML}{018571}
\definecolor{marroon}{HTML}{881c1c}

\usepackage{mathtools}

\usepackage{amsmath}
\usepackage{array}
\usepackage{xcolor}
\usepackage{arydshln}
\usepackage{siunitx}
\colorlet{tablerowcolor4}{gray!50} 

\newcommand*{\textlabel}[2]{%
  \edef\@currentlabel{#1}
  \phantomsection
  #1\label{#2}
}
\usepackage{tcolorbox}

\colorlet{tableheadcolor}{gray!25} 
\colorlet{tablerowcolor}{gray!15} 
\colorlet{tablerowcolor2}{gray!45} 
\colorlet{tablerowcolor3}{gray!25} 

\newcommand{\rowcollight}{\rowcolor{LightGray}} %

\newif{\ifhidecomments}
  \hidecommentsfalse 
\ifhidecomments
    \newcommand{\melissa}[1]{}
    \newcommand{\dongwhi}[1]{}
    \newcommand{\koustuv}[1]{}
    \newcommand{\violeta}[1]{}
\else
    
    \newcommand{\melissa}[1]{\textbf{\small\sffamily{\textcolor{dolive}{[#1 -- Melissa]}}}}
    \newcommand{\dongwhi}[1]{\textbf{\small\sffamily{\textcolor{dpink}{[#1 -- Dong Whi]}}}}
    \newcommand{\koustuv}[1]{\textbf{\small\sffamily{\textcolor{violet}{[#1 -- Koustuv]}}}}
  \fi
\newcommand{\edit}[1]{{\textcolor{editCol}{#1}}}

\colorlet{tableheadcolor}{gray!25} 
\colorlet{tablerowcolor}{gray!5} 

\definecolor{neutralCol}{HTML}{dd1c77}
\definecolor{neutralGreen}{HTML}{31a354}
\definecolor{NewBlue}{HTML}{1879ba}
\definecolor{bleudefrance}{rgb}{0.19, 0.55, 0.91}  
\definecolor{AfTrColor}{HTML}{0868ac}  
\definecolor{BfTrColor}{HTML}{a8ddb5}  

\definecolor{AfCtColor}{HTML}{b10026}  
\definecolor{BfCtColor}{HTML}{fd8d3c}

\graphicspath{ {figures/} }

\renewcommand{\textcolor}[2]{#2}

\AtBeginDocument{%
  \providecommand\BibTeX{{%
    \normalfont B\kern-0.5em{\scshape i\kern-0.25em b}\kern-0.8em\TeX}}}

\setcopyright{cc}
\setcctype{by}
\acmJournal{PACMHCI}
\acmYear{2026} \acmVolume{10} \acmNumber{6} \acmArticle{CSCW110}
\acmMonth{10} \acmDOI{10.1145/3816958}

\begin{document}

\title[\edit{Journeys of Parents with LGBTQ+ Children}]{\edit{Journeys of Parents with LGBTQ+ Children: How Trauma and Healing Reshape Identity and (Mis)Informating Practices}}


\author{Soonho Kwon}
\email{soonho@gatech.edu}
\orcid{0000-0002-2783-6364}
\affiliation{%
  \institution{Georgia Institute of Technology}
  \city{Atlanta}
  \state{GA}
  \country{USA}
  \postcode{30332}}

\author{Dong Whi Yoo}
\email{dy22@iu.edu}
\orcid{0000-0003-2738-1096}
\affiliation{%
  \institution{Indiana University Indianapolis}
  \city{Indianapolis}
  \state{IN}
  \country{USA}
  \postcode{46202}}

\author{Koustuv Saha}
\email{ksaha2@illinois.edu}
\orcid{0000-0002-8872-2934}
\affiliation{%
  \institution{University of Illinois Urbana-Champaign}
  \city{Champaign}
  \state{IL}
  \country{USA}
  \postcode{61801}}

\author{Shaowen Bardzell}
\email{shaowen@cc.gatech.edu}
\orcid{0000-0001-7596-9244}
\affiliation{%
  \institution{Georgia Institute of Technology}
  \city{Atlanta}
  \state{GA}
  \country{USA}
  \postcode{30332}}

\author{Younah Kang}
 \email{yakang@yonsei.ac.kr}
 \orcid{0000-0003-4981-6329}
 \affiliation{%
   \institution{Yonsei University}
   \city{Seoul}
   \country{Republic of Korea}
   \postcode{03722}}

\renewcommand{\shortauthors}{Kwon et al.}

\begin{abstract}

This study examines how parents of LGBTQ+ individuals in South Korea navigate the emotional rupture fueled by fear, isolation, and disorientation after learning their children’s queer identity, encounter queer-related (mis)information as a way of coping with this emotional toll, and come to listen to queer realities relationally. Through this process, we highlight how parents reconstruct their identities as supportive parents, which reshapes their informating practices, making them more critical in assessing queer-related (mis)information, developing strategies to protect themselves from harmful narratives, and actively challenging misinformation to support others navigating similar experiences.

This work contributes to CSCW by (1) foregrounding parents of LGBTQ+ individuals, an underrepresented yet critical stakeholder group in Queer HCI; (2) demonstrating how identity reconfiguration following a trauma-healing process could transform informating practices; and (3) arguing that addressing misinformation requires attention beyond individual fact-based discerning to account for its relational, cultural, and emotional dimensions. Further, we invite CSCW scholars to reconsider the balance between abstracting and humanizing information, explore future design possibilities for parents of LGBTQ+ children, and reflect on the role of researchers as participants in collective research communities fueled by care.

\end{abstract}




\maketitle

\section{Introduction}
Parental relationships play a pivotal role in shaping LGBTQ+\footnote{The terminology of \textit{LGBTQ+} and \textit{queer} in HCI has not reached a fixed consensus, as discussed by \citeauthor{taylor2024cruising} \cite{taylor2024cruising}. Broadly, we observe how LGBTQ+ tends to refer to individuals, while queer is more often used to describe non-normative methods or perspectives—though defining either term too rigidly may run counter to the ethos of queerness itself. In the Korean context, terms such as LGBTQ+, queer, and sexual minority are often used interchangeably, with a focus on inclusive, collective identities that strengthen solidarity among queer communities \cite{jeong2024moving}. Accordingly, we use both terms as interchangeable umbrella terms rather than strictly defining them. Notably, we refer to Parents, Families and Allies of LGBTAIQ+ People in Korea (PFLAG Korea) using their official name, which includes the word LGBTAIQ+.} individuals' well-being. A lack of familial support—such as denial, rejection, or abuse—has been identified as one of the most harmful factors impacting LGBTQ+ well-being~\cite{richter2017examining,grossman2021parental,roe2017family}, while affirming support from parents is strongly associated with improved mental health, greater resilience, and enhanced quality of life~\cite{ryan2010family,needham2010sexual,homma2007emotional}. These findings position parents as critical stakeholders for the queer HCI and CSCW studies, particularly in addressing the sociotechnical dimensions of queer well-being.

Despite its importance, accepting their children's\footnote{In this study, we use the term child to refer to an offspring, regardless of age, rather than a young person below the age of puberty or legal adulthood.} queer identity is often a complex and emotionally tolling experience, especially for parents raised in cultures where queerness is invisible, misunderstood, or even demonized. In the Korean context, parents learning their children's identity often leads to significant emotional distress—described by many as traumatic—leaving them uncertain about how to respond, fearful for their children's well-being, and isolated due to the lack of guidance on moving forward \cite{pflag}. During these moments, misinformation about queer realities frequently compound the problem. Such misinformation offered false assurances (e.g., claims that conversion therapy can “cure homosexuality”) or amplified catastrophic fears (e.g., health misinformation suggesting that male-to-male sexual contact inevitably leads to HIV/AIDS).

In our study, we trace the journeys of ten Korean parents of LGBTQ+ children as they navigate the emotional experience of learning about their children's identities. We showcase how parents' initial distress largely stems from the feeling of \textit{fear} rooted in concern for their children's well-being and future, \textit{disorientation} on what the next step is to understand and help their children, and \textit{isolation} of having no one to confide in about these experiences. We further examine how parents seek diverse informational sources to recover from such distress and eventually reshape their identity as a supportive parent, ultimately leading to changed informating practices. By doing so, our work builds on CSCW scholarship that highlights how identities are reconstructed through trauma-healing journeys \cite{randazzo2025kintsugi}, frequently with the support of what prior work describes as “human infrastructures”—individuals who provide emotional and informational support to help mitigate both emotional and practical concerns \cite{ammari2025finding, zehrung2025did}. Ultimately, we demonstrate how this reconstructed sense of parental identity, shaped through a trauma-healing journey, influences parents' information-seeking and validation practices.

In doing so, we examine misinformation as a key phenomenon. \textcolor{blue}{Specifically, we present misinformation as a `graspable', yet ineffective, solution to recover from isolation and disorientation, and showcase how parents' recovery journey changes their informating practices to dispel, protect themselves against, and even actively fight back on misinformation.} Earlier misinformation studies focused on tools that help users assess credibility—such as source tracking~\cite{chen2022visualbubble, lee2022misvis, Jahanbakhsh2021exploring} or interface designs that flag misleading content~\cite{jahanbakhsh2023exploring, kirchner2020countering}. Yet, recent work highlights that misinformation persists not because people lack reasoning skills, but because of emotional and relational factors such as affirming social bonds or emotional resonance with content~\cite{allen2022birds,varanasi2022accost,efstratiou2022adherence,hassoun2023practicing,huang2015connected}. Building on this critique, scholars have been emphasizing that navigating misinformation should not be treated merely as an individual's cognitive task but one that requires situated understandings accounting for the societal, relational, and emotional dynamics at play \cite{haque2020combating, sultana2021dissemination}. We contribute to this conversation by examining the role of misinformation as an attractive means of mitigating emotional trauma and how dispelling it requires attending to social and relational conditions.

To unpack our empirical findings, we draw on the concept of \textit{listening} from care ethics \cite{gilligan1993different, noddings2013caring} as an analytic lens. We use \textit{listening}—the intentional act of setting aside one's worldview to deeply resonate with those in vulnerable positions—to identify and organize key moments in parents' journeys. \textcolor{blue}{Our findings show that \textit{listening} to queer realities plays a crucial role in parents' recovery journeys, contributing to the reconstruction of their identities as supportive figures and ultimately reshaping their informating practices.}

Our work contributes to CSCW research by (1) positioning parents of LGBTQ+ individuals as critical stakeholders in the sociotechnical dynamics of queer experience, (2) illustrating how parents' information-seeking and validation practices transform through a relational and emotionally grounded healing process supported by human infrastructures; and (3) demonstrating that addressing misinformation goes beyond purely rational fact-checking and detection, instead requiring more context-sensitive and value-driven approaches.

\textit{\textbf{Disclaimer}: In reviewing this paper alongside multiple LGBTQ+ individuals, we recognized the potential emotional toll this paper may carry, particularly for those who have experienced strained or painful relationships with their families. As a team led by a queer researcher, we clearly state that this paper does not seek to justify or excuse any harm caused by families to LGBTQ+ individuals. We acknowledge that the supportive family case that we present in this paper may feel distant or unattainable for many. We stand in solidarity with all those who have endured hardship in navigating family relationships and affirm their experiences with care and respect.}

\section{Related Work}
\subsection{Parents of LGBTQ+ Individuals}

Queer HCI encompasses research that (1) is conducted by LGBTQ+ researchers, (2) centers LGBTQ+ populations within HCI discussions, or (3) employs queer approaches as methodologies to critically interrogate the field~\cite{devito2021queer, taylor2024cruising}. One critical yet underexplored dimension in the field is the role of the surrounding stakeholders of queer individuals, particularly parents. While some HCI research has examined how parents use digital tools to support their LGBT+ children, such as engaging in social VR communities to connect them with peer support~\cite{acena2021in}, this remains a relatively limited area of study. 

Our study seeks to address and expand this conversation by foregrounding parents' role in shaping LGBTQ+ individuals' well-being, an area we believe deserves deeper recognition within CSCW and HCI. Below, we explore the significance of parental support in queer well-being, along with an overview of parent–child dynamics in our research site, South Korea.

\subsubsection{Parental Support}
For LGBTQ+ adolescents, parental support is a significant protective factor~\cite{ryan2010family}, helping reduce depression, suicidal thoughts, and substance use, while promoting emotional well-being~\cite{needham2010sexual,homma2007emotional}. Unfortunately, many LGBTQ+ youths lack supportive parental relationships, which can become a major source of distress and, in extreme cases, an issue of survival when they are forced out of their homes~\cite{haas2010suicide, king2008systematic}. Studies show that nearly 70\% of queer youth face parental rejection of their sexual identity, with this being particularly common in certain cultural communities~\cite{richter2017examining}. Additionally, nearly half of transgender and gender-nonconforming youth experience minority stress due to parental reactions~\cite{grossman2021parental}. Even when parents offer support, their lack of understanding of queer realities may prevent it from fully aligning with their child's needs~\cite{roe2017family}. That said, we position parents and close family members as critical stakeholders who, if better informed, can play a vital role in promoting the well-being of LGBTQ+ individuals.

\subsubsection{Queer in Korea}
In examining the nuanced parental relationships of queer individuals, it is crucial to situate our study within the South Korean context, one of the few OECD nations that actively persecutes queer individuals~\cite{globalbarometers,Mitsanas}. Korea's hostility toward LGBTQ+ individuals is visible in both political discourse and public attitudes: major presidential candidates from both major parties denounce homosexuality, state officials spread misinformation linking LGBTQ+ individuals to HIV/AIDS or communism~\cite{Kuusisto_2023,CBCnews_2017,park2024}, and surveys show low public acceptance of queer identities and same-sex marriage~\cite{Gallup_2023,So-Yeon_2023}. Such social atmospheres extend to the systematic discriminations: in the military, where all non-disabled men are required to serve, criminalizes consensual same-sex relations and expels transgender personnel~\cite{Mitsanas,Kwaak_2017,ABCNews_2021}; the national curriculum recently excluded all LGBTQ+ related content, while schools often lack adequate mental health support for queer students~\cite{Human2021}. Against this backdrop, LGBTQ+ youth in Korea face alarmingly high rates of suicidal ideation and attempts~\cite{kang2006}.

In this hostile environment, many Korean parents react negatively when their children come out, shaped by internalized homophobia, misconceptions about queer identities, and fears for their children's safety~\cite{pflag}. A recent study of 2,381 Korean LGBs found that disclosure to family was more often a source of stress and depressive symptoms than liberation~\cite{lee2024gender}. In response, several organizations have worked to support queer rights. Of particular relevance to our study is PFLAG Korea (Parents, Families, and Allies of LGBTAIQ+ in Korea), which has supported queer individuals and their families since 2014. 

Against this backdrop, we position our work within the ongoing effort to expand non-Western queer representation in CSCW and HCI \cite{park2025stuck}. Recognizing that queer experiences in Asian contexts often differ from those in the West, such as through the emphasis on family honor, goal of staying discrete and safe rather than being ``out and proud'', or the subtler manifestations of homophobia \cite{laurent2005sexuality,park2025stuck}, we do not seek to generalize our findings to the global queer population. Rather, we offer them as a situated, local queer account that brings an underrepresented reality into the CSCW community, encouraging further consideration of how our work could better attend to culturally nuanced queer experiences.

We also disclose that this study was conducted with the help of PFLAG Korea. The organization provided support for participant recruitment, and the leading members helped in shaping our advocacy-oriented study goals \cite{harrington2019deconstructing}.

\subsection{Understanding Parental Transformation Through Trauma-Informed Frameworks}

We frame the parental journeys of learning and accepting their children's queer identities as processes of recovering from emotionally traumatic experiences. This framing is in no way meant to suggest that having an LGBTQ+ child is inherently traumatic. Rather, drawing from prior literature by PFLAG Korea \cite{pflag} and our engagement with participants, our goal is to acknowledge the emotional distress many parents described after learning of their children's queer identities—particularly in the Korean context where queerness is deeply stigmatized. These conditions often lead to parents' concerns for their children's safety and well-being, feelings of isolation where they have no one to discuss these matters, as well as disorientation about future steps following identity disclosure. In such contexts, we highlight how queer identity disclosure can be traumatic not only for the queer individual but also in intergenerational and family-oriented ways, especially within Asian cultural frameworks where expectations around filial piety and saving face carry significant weight \cite{laurent2005sexuality}.

Trauma-informed scholarship emphasizes the importance of designing, discussing, and managing technologies with an awareness of the trauma carried by individuals, communities, and societies—prioritizing user safety, well-being, and, most relevant to our work, healing and recovery \cite{randazzo2023trauma,chen2022trauma}. Online spaces, as sites of both trauma-inducing misinformation and healing, have been central to this approach \cite{scott2023trauma}. \citeauthor{ammari2025finding} and \citeauthor{zehrung2025did} highlight how human infrastructures, or people who share similar experiences, play a crucial role by offering emotional and informational support that helps individuals navigate and recover from trauma \cite{ammari2025finding, zehrung2025did}. Such engagements often take the form of information activism or collective meaning-making, which CSCW scholars have examined as core mechanisms of recovery in marginalized communities \cite{jonas2024better,saha2019language}. Following recovery, \citeauthor{randazzo2025kintsugi} describe how individuals often reconstruct their identities through these relational and communal processes \cite{randazzo2025kintsugi}.

Our work builds upon this rich scholarship by illustrating how Korean parents, through collaborative sensemaking with the broader queer community rooted in \textit{listening}, reconstruct their identities as supportive parents after the emotional rupture of learning their children's identities. We further expand this scholarship by showing how this reconstructed identity meaningfully transforms their information-seeking and validation practices related to their traumatic experiences.

\subsection{Misinformation as Collaborative Sensemaking in CSCW}
In our work, we present misinformation as a key compounding factor that fuels the emotional distress experienced by parents. Misinformation causes a wide range of harms, from inciting hatred towards minority groups and fostering hostility~\cite{scott2023exploring}, to spreading inaccurate health information to vulnerable populations, such as transgender individuals~\cite{Augustaitis2021online,garofalo2023editor}. For the LGBTQ+ community, the selective amplification of misinformation through social and public media fuels stigma, discrimination, and anti-LGBTQ+ policies, harming mental health and well-being~\cite{Augustaitis2021online, Kremidas-Courtney2023, Watters2024pride}.
 
Prior studies on misinformation proposed strategies to aid users' conscious evaluation of content, including presenting intuitive source information~\cite{urakami2022finding,heuer2022comparative}, designing tools for critical assessment~\cite{chen2022visualbubble,lee2022misvis,Jahanbakhsh2021exploring,dingler2021workshop}, developing flagging interfaces~\cite{jahanbakhsh2023exploring,kirchner2020countering}, and implementing AI-based credibility checks~\cite{lu2022effects}. More recent CSCW research, however, highlights the social dynamics behind misinformation, rather than treating it as an isolated phenomenon~\cite{aghajari2023reviewing}. Scholars highlight how people share misinformation not because they believe it to be accurate, but because it aligns with their values or sustains social bonds~\cite{allen2022birds,varanasi2022accost,efstratiou2022adherence,hassoun2023practicing,huang2015connected}. These dynamics are evident in work showing how health misinformation is entangled with class in India~\cite{varanasi2022accost} or with local belief systems in Bangladesh~\cite{sultana2021dissemination}.

Resonating with this perspective, CSCW scholars argue that navigating misinformation should be understood as a collaborative, social process embedded within information infrastructures \cite{haque2020combating, sultana2021dissemination, juneja2022human}. \citeauthor{chen2024we}, for instance, show how “trusted messengers” within Black and Latinx communities played crucial roles in combating vaccine misinformation, where human infrastructure, rather than a technical one, is a key consideration in understanding misinformation dynamics \cite{chen2024we}. Such a holistic understanding of misinformation dynamics becomes even more important in close relationships \cite{vraga2017using,vraga2020correction}. Correcting misinformation is especially fraught among family members, where it intersects with hierarchies, religion, sexism, and language differences~\cite{martinez2025would}. Addressing misinformation, therefore, requires attention to social and emotional dynamics \cite{urakami2022finding}, highlighting the need for holistic understandings that move beyond purely design-solution approaches. 

Our work builds on these works by presenting a rich empirical account that showcases the relational and emotional characteristics of misinformation, both in how misinformation proliferates through parents' relational concerns for their children, and in how it is dispelled through parents' reconstructed identities as supportive caregivers after a relational healing journey, rather than those solely based on cognitive and rational bases.

\subsection{Care Ethics and the Act of \textit{Listening}}
In our study, we borrow the concept of \textit{listening} from care ethics as a lens to unpack our empirical findings. In care ethics, \textit{listening} is defined as moments where those in a position of power lower their defenses, set aside their worldviews and prejudices, and actively engage with those who are less powerful, seeking to understand their perspectives. \textit{listening} is an intentional act where the person in power seeks to grasp the marginalized's position, emotions, and perspective without preconceived judgments, ultimately fostering trust and empathy~\cite{gilligan2014moral}.

\citeauthor{gilligan2014moral} draws on the work of \citeauthor{shay2010achilles}, who studied the treatment of Vietnam veterans with post-traumatic stress disorder (PTSD), to illustrate this concept \cite{gilligan2014moral}. Shay discusses how, in many clinical settings, professionals often hear their patients through the lens of their own worldview, categorizing or diagnosing experiences based on pre-existing frameworks or judgments. Shay points out that this could erode trust by prioritizing the professional's interpretation over the patient's lived experience. He advocates for \textit{listening} that immerses oneself fully in another's narrative to avoid such interactions~\cite{shay2010achilles}. The act of \textit{listening} is further highlighted in many subsequent studies, highlighting the importance of attentiveness and engagement \cite{robinson2011stop, tronto2020moral}

We chose \textit{listening} as a lens to unpack our empirical findings because it provides a powerful analytic frame for understanding how parents move from emotional rupture to relational repair. In unpacking our empirical findings, we identify and attend to moments where parents set aside preconceived notions about queerness (often fueled by prevailing misinformation), lower their guard, and begin \textit{listening} to their children and broader queer realities. Through this process, we show how \textit{listening} cultivates empathy toward LGBTQ+ individuals, initiates recovery from initial emotional trauma rooted in fear toward queer realities, and establishes a relational foundation for reconstructing their identities as supportive parents.

\section{Study Design}
\subsection{Participants}
To conduct our study, we recruited parents of LGBTQ+ individuals who are supportive of their identities. Participants were recruited through PFLAG Korea, university LGBTQ+ communities, researchers' networks, and snowball sampling. In total, ten parents participated. Table \ref{tab:participants} summarizes demographics, including the child's identity, age, and years since learning their identities.

Despite extensive recruitment efforts over several months, we acknowledge that finding participants was deeply challenging. This difficulty likely reflects the fact that, sadly, only a handful of parents in South Korea both support their LGBTQ+ children and are willing to speak openly about their experiences, mirroring the country's broader social atmosphere. While a “sample” of 10 participants might be insufficient for “saturation”, we position our study on epistemological and methodological grounds within the feminist and action research traditions in HCI and CSCW. These traditions emphasize understanding qualitative accounts as windows into underrepresented realities rather than as data for broad generalization \cite{bardzell2010feminist} and in search for “transferability” rather than “generalizability” \cite{hayes2011relationship}.

\begin{table}[h]
\centering
\caption{Participant Demographics}
\sffamily
\small

\begin{tabular}{ccccccccc}
& \textbf{Age} & \textbf{Gender} &  \textbf{QC Identity} & \textbf{QC Age} & \textbf{Years Known} \\
\toprule
P1 & 64 & Female &  MTF  & 29 & 8 \\
\rowcollight P2 & 48 & Female & MTF & 15 & 4 months \\
P3 & 68 & Female &  Gay & 41 & 17 \\
\rowcollight P4 & 69 & Female &  Lesbian → FTX &  38 & 24 → 4 \\
P5 & 63 & Female &  Gay & 31 & 10 \\
\rowcollight P6 & 58 & Female &  Gay & 29 & 3 \\
P7 & 60 & Female & Gay & 32 & 17 \\
\rowcollight P8 & 52 & Female & MTF, Lesbian & 27 & 8 \\
P9 & 59 & Male & Gay & 32 & 17 \\
\rowcollight P10 & 52 & Female & Lesbian & 27 & 9 \\
\Description[table]{This table provides demographic details of 10 participants, labeled P1 through P10. It includes their age, gender, number of children, the birth order of their queer child (QC), the queer child's identity, the queer child's current age, and the number of years the participant has known about their child's queer identity. P1 is a 64-year-old female with two children, whose second-born child is a 29-year-old MTF (Male-to-Female) who has known their identity for 8 years. P2 is a 48-year-old female with two children, whose first-born is a 15-year-old MTF, and the participant has known for 4 months. P3 is a 68-year-old female with two children, whose second-born is a 41-year-old gay male, and the participant has known for 17 years. P4 is a 69-year-old female with three children, whose second-born child is a 38-year-old lesbian who transitioned to FTX, and the participant has known about their lesbian identity for 24 years and about the transition for 4 years. P5 is a 63-year-old female with two children, whose second-born is a 31-year-old gay male, and the participant has known for 10 years. P6 is a 58-year-old female with two children, whose second-born is a 29-year-old gay male, and the participant has known for 3 years. P7 is a 60-year-old female with two children, whose first-born is a 32-year-old gay male, and the participant has known for 17 years. P8 is a 52-year-old female with two children, whose first-born is a 27-year-old MTF and lesbian, and the participant has known for 8 years. P9 is a 59-year-old male with two children, whose first-born is a 32-year-old gay male, and the participant has known for 17 years. P10 is a 52-year-old female with two children, whose first-born is a 27-year-old lesbian, and the participant has known for 9 years.}

\end{tabular}
\label{tab:participants}
\end{table}

\subsection{Study Procedure}
We conducted semi-structured interviews (60–80 minutes) in person or online. Participants received a 30,000 KRW (≈21.67 USD) gift card after the interview, with one donating theirs to PFLAG Korea. All interviews were led by the first author, with some attended by the second author, who occasionally asked follow-up questions. With consent, interviews were recorded, transcribed, anonymized during transcription, and shared among the authors. Below is a detailed outline of the interview sessions. 

We began the interview by introducing the study procedure and the concept of misinformation to ensure participants understood our definition. Participants were given a handout explaining the concept and types of misinformation (see Table \ref{tab:misinformation_categories})~\cite{Iacucci_2021}. This ensured that participants provided more informed, contextually relevant responses to misinformation. Before moving on to the interview, they were encouraged to review the handout and ask any clarifying questions.

\renewcommand{\arraystretch}{1.3} 
\begin{table}[h]
\centering
\caption{Categories of Misinformation \cite{Iacucci_2021}}
\sffamily
\small
\begin{tabular}{p{0.17\linewidth}p{0.36\linewidth}p{0.36\linewidth}}

\textbf{Type}& \textbf{Explanation} & \textbf{Examples} \\ \toprule
Fabricated Content & Content that is completely false. & A completely false news article claiming that a celebrity has passed away when they haven't\\ 
\rowcollight Manipulated Content & Content that distorts genuine information or images. & Adding specific elements to an image through Photoshop.\\ 
Imposter Content & Content that pretends to be from a genuine source. & Someone posing as a celebrity on social media to express political opinions.\\ 
\rowcollight Misleading Content & Content that misleads or causes misunderstanding. & Stating an opinion as fact, like "This movie is the best ever made."; Misreporting only part of a speech, distorting the speaker's intent.\\ 
False Context & Situations where accurate data or materials are used in an incorrect context. & Using a photo of a tsunami in Japan with a headline "Tsunami Hits South Korea."\\ 
\rowcollight Satire and Parody & Content that conveys false information with humorous intent. While not malicious, it can lead to misinformation. & A humorous post about the Earth being flat being reported as a real article.\\
False Connections & When the title, visuals, or descriptions are disconnected from the actual content. & Sensational titles that do not match the content to attract clicks.\\ 
\rowcollight Sponsored Content & Advertisements disguised as regular content. & An article that appears to be a news report but is actually an advertisement for a new product\\ 
Propaganda & Content designed to instill certain values, attitudes, or knowledge. & A video produced to promote a political agenda by spreading exaggerated claims about the opposition\\ 
\rowcollight Error & Mistakes made by media organizations during reporting. & A news outlet mistakenly reporting incorrect election results due to a typo in their data\\ 
\bottomrule
\Description[table]{This table outlines different categories of misinformation, their explanations, and examples. It lists 10 types of misinformation. Fabricated Content refers to entirely false content, such as a fake news article about a celebrity's death. Manipulated Content involves distorting genuine information, like adding elements to an image using Photoshop. Imposter Content pretends to come from a legitimate source, such as someone impersonating a celebrity on social media. Misleading Content causes misunderstandings, such as presenting an opinion as fact. False Context uses accurate information but in an inappropriate context, such as using a photo from one disaster for another event. Satire and Parody involve humorous but potentially misleading content, such as a joke article on the Earth being flat. False Connections occur when titles or visuals are unrelated to the actual content, often for clickbait. Sponsored Content disguises ads as regular articles. Propaganda is designed to promote a specific agenda, while Error refers to unintentional mistakes in reporting, such as a media outlet reporting incorrect election results due to a typo.}

\end{tabular}

\label{tab:misinformation_categories}
\end{table}
\renewcommand{\arraystretch}{1}

First, we asked how participants learned about and accepted their children's identities, highlighting the information sources they consulted and their experiences in navigating this information. Then, we conducted an exercise where participants were presented with articles containing misinformation commonly found on Korean search engines like Google and Naver\footnote{Korea's top search engine}. We provided participants with five articles containing misinformation, selected through the following process: (1) We searched the words gay, lesbian, homosexual, bisexual, transgender, and queer, along with keywords commonly associated with LGBTQ+ misinformation (e.g., AIDS/HIV, choice, conversion therapy) on Google and Naver. (2) We collected articles (including webpages, news, and blog posts) from the first five pages of each search. (3) We identified misinformation articles based on the United Nations criteria for misinformation~\cite{Iacucci_2021}. (4) We categorized the articles by topic and named each category. (5) We selected one or two articles per category, considering length, clarity, and relevance to our topic, ensuring a balanced representation of issues related to both sexual orientation (e.g., gay, lesbian, bisexual) and gender identity (e.g., transgender, non-binary). We ended up with four categories and five articles, as shown in \ref{tab:misinformation_categories}, with translated articles in the appendix. We presented these articles and asked participants whether they had encountered or engaged with similar content, whether they considered it to be misinformation, how they (would have) reacted at the time, and how they respond to similar articles now.

Finally, we asked about their experience in encountering other queer-related misinformation, the contexts in which it occurred, their initial responses, how they recognized it as misinformation, and how they reacted to it. We further inquired about broader responses related to the issue, including how their views, attitudes, and practices toward queer-related information have changed after accepting their children's identities.

\subsection{Ethical Considerations}
In designing these sessions, along with the approval from the university ethics review board, we took special care to address the potential emotional toll on participants and the risk of exploitation \cite{le2015strangers,harrington2019deconstructing}. Given the personal and sensitive nature of the topic, we prioritized participants' emotional safety while also remaining mindful of avoiding exploitation, especially in light of a case shared by PFLAG Korea in which a researcher misrepresented the group and portrayed participants recruited from it in an adversarial tone.

To uphold these commitments, we consulted with PFLAG Korea to ensure our materials were appropriate before conducting the interview. The first author also disclosed his identity as a gay man, both in consultations with PFLAG and in each interview, to share his personal motivation for this research and explicitly show how we advocate for queer individuals and their families. Finally, we sent the semi-final draft of this paper to both PFLAG and participants and revised the final version based on their feedback to ensure their narratives were accurately and respectfully represented.

\renewcommand{\arraystretch}{1.5} 
\begin{table}[h]
\centering

\caption{Misinformation Articles on Homosexuality and Gender Issues}
\sffamily
\small
\begin{tabular}{p{0.31\linewidth}p{0.45\linewidth}p{0.15\linewidth}}

\textbf{Category} & \textbf{Article Title} & \textbf{Source} \\ \toprule
Homosexuality Causes Diseases & \textbullet~Homosexuality is a Cause of Diseases - Homosexual Acts Are the Main Cause of AIDS... Helping People Quit Homosexuality is True Love & GBN News \\ 
\rowcollight Sex Reassignment is Harmful to Health & \textbullet~Sex Reassignment Surgery Leads to Lifelong Suffering... Suicide Rate is 22 Times Higher & The Mission News \& Kukmin Daily \\ 
LGBTQ Individuals Live Unhappy Lives & \textbullet~Homosexuality is Not Love. They Grow Old Alone and Eventually End Up Miserable: Testimony of Korea's First Transgender Person Kim Yu-bok on Their Deathbed & Kukmin Daily \\ 
\rowcollight Homosexuality is Acquired and Treatable & \textbullet~Case Studies of Homosexuality Treatment & Credo Association \\
\rowcollight & \textbullet~``Homosexuality Treatment Research Shows 79\% Average Effectiveness''... Overturning the ``Innate'' Claim & Kukmin Daily \\ 
 \bottomrule
 \Description[table]{The table titled "Misinformation Articles on Homosexuality and Gender Issues" presents four categories of misinformation along with related article titles and sources. The first category, "Homosexuality Causes Diseases," includes the article "Homosexuality is a Cause of Diseases - Homosexual Acts Are the Main Cause of AIDS... Helping People Quit Homosexuality is True Love," sourced from GBN News. The second category, "Sex Reassignment is Harmful to Health," features the article "Sex Reassignment Surgery Leads to Lifelong Suffering... Suicide Rate is 22 Times Higher," sourced from The Mission News and Kukmin Daily. The third category, "LGBTQ Individuals Live Unhappy Lives," includes the article "Homosexuality is Not Love. They Grow Old Alone and Eventually End Up Miserable: Testimony of Korea's First Transgender Person Kim Yu-bok on Their Deathbed," sourced from Kukmin Daily. The final category, "Homosexuality is Acquired and Treatable," includes the articles "Case Studies of Homosexuality Treatment" and "'Homosexuality Treatment Research Shows 79\% Average Effectiveness'... Overturning the 'Innate' Claim," sourced from Credo Association and Kukmin Daily.}
\end{tabular}

\label{tab:misinformation_articles}
\Description[table]{This table lists articles that spread misinformation related to homosexuality and gender issues, categorized by topic. Under the category "Homosexuality Causes Diseases," an article titled "Homosexuality is a Cause of Diseases - Homosexual Acts Are the Main Cause of AIDS... Helping People Quit Homosexuality is True Love" was published by GBN News. In the category "Sex Reassignment is Harmful to Health," an article titled "Sex Reassignment Surgery Leads to Lifelong Suffering... Suicide Rate is 22 Times Higher" was published by The Mission News and Kukmin Daily. The category "LGBTQ Individuals Live Unhappy Lives" includes the article "Homosexuality is Not Love. They Grow Old Alone and Eventually End Up Miserable: Testimony of Korea's First Transgender Person Kim Yu-bok on Their Deathbed" from Kukmin Daily. Lastly, under "Homosexuality is Acquired and Treatable," two articles are listed: "Case Studies of Homosexuality Treatment" and "Homosexuality Treatment Research Shows 79\% Average Effectiveness... Overturning the 'Innate' Claim," both published by Credo Association and Kukmin Daily.}

\end{table}
\renewcommand{\arraystretch}{1}

\subsection{Analysis}

Based on the transcripts, the first and second authors conducted a reflexive thematic analysis (RTA). RTA foregrounds researchers' perspectives in interpreting data to provide a situated understanding of complex social contexts, rather than assuming neutrality or objectivity~\cite{Braun2021thematic}. In RTA, many practices emphasized in traditional thematic analysis—such as linear, structured coding procedures or the production of formal codebooks—are de-emphasized or even discouraged, as they can dilute the researcher's interpretive agency and imply an undue sense of generalization about specific codes. RTA instead prioritizes a nonlinear, iterative, and embodied analytic engagement by individual researchers, where familiarization with the data, coding, and theme development often overlap, and analytic insights emerge through these fluid and messy processes. Following this spirit, our analysis sought to develop narratives that felt true to and resonant with parents' realities as we came to understand and connect with them through the interviews, rather than focusing on a rigid analytical framework. Nevertheless, we present our analytic process as faithfully and transparently as possible below.

\textcolor{blue}{Following the stages of RTA~\cite{Braun2021thematic}, the first and second authors began by familiarizing themselves with the data. This allowed us to identify five key stages of parents' journeys: (1) before learning their child's identity, (2) the moment of discovering their child's identity leading to significant emotional distress, fueled by disorientation and isolation, (3) seeking and encountering diverse queer-related (mis)information to recover from such distress, (4) reconstructing their identities as supportive parents after interacting with queer realities more directly, and (5) subsequently transforming their information-seeking and validation practices. Taking a deductive approach, we assigned semantic codes to quotes that corresponded to these phases.}

 \textcolor{blue}{Then, using a bottom-up inductive approach, we applied latent codes to particularly insightful remarks, allowing us to surface implicit meanings and underlying implications within participants' narratives. Through this inductive process, we observed recurring patterns in which parents described letting go of prior worldviews and engaging with queer realities in ways that facilitated their recovery. We found that these patterns resonated with the notion of \textit{listening} as articulated in care ethics. We therefore draw on \textit{listening} as an interpretive lens to revisit and make sense of our empirical findings, particularly in relation to parents' identity reconstruction and subsequent changes in how they engage with, respond to, and manage misinformation.}

\subsection{Positionality Statement}
The authors' positionalities within the Korean queer context deeply shaped the framing and presentation of this research. As a gay man from Korea, the first author's lived experience, particularly his experience of disclosing his identity to his parents and navigating the subsequent journey, deeply informed and influenced the analytical and interpretive processes. Drawing on this perspective and his ongoing engagement with LGBTQ+ advocacy, the first author led the interview sessions and the data analysis.

As a team, all authors identify as East Asian or South Asian, with three originally from South Korea. We are currently based in either South Korea or the U.S., with the first author traveling from Korea to the U.S. institute during the course of this project. We acknowledge that our academic affiliations and disciplinary communities provided us with the privilege and support to engage openly with queer realities. 

All authors contributed their expertise in HCI research on LGBTQ+ communities, mental health, feminist theory, and work with marginalized populations, offering critical guidance throughout the project's development.

\section{Findings}
\subsection{Before Learning the Child's Identity: Information Vacuum and Invisibility}

Before discovering their child's identity, many parents viewed queer issues as distant and irrelevant to their lives. This led them to consume queer-related information passively without much reflection. As one participant shared, \textit{``I didn't react (to queer-related information) much because I thought it wasn't my issue. I only understood it through the information given to me, and since it didn't concern me, I didn't actively seek out more'' (P6)}

Media portrayals further fueled their distorted understandings of LGBTQ+ individuals with negative or mocking depictions. One participant recalled, \textit{``The memories that remained with me were of the media that made fun of sexual minorities. Because of that, I think I also developed some prejudices'' (P3). }

This passive stance often resulted in misunderstandings and biases. For instance, some parents confused different identities, such as equating being gay with being transgender. This phase reflects a state where parents did not see the need to \textit{listen} to queer voices, neglecting their stories and realities due to the belief that these issues were unrelated to them.

\subsection{Learning the Child's Identity: Crisis and Rupture}
\subsubsection{Parental Emotional Trauma}
Discovering their child's queer identity, whether through self-disclosure or other means—such as being outed by the child's friends or finding a letter exchanged with a romantic partner—was often a deeply shocking experience for parents. Even among parents who were not overtly negative or hostile toward LGBTQ+ people, moments of reassurance and expressions of love toward their children often coexisted with a flood of mixed emotions, particularly fear arising from uncertainty about what it means to be queer and how to navigate what lies ahead. A pervasive feeling among parents was an overwhelming sense of isolation, often described as \textit{``becoming a minority together.''} One parent reflected, 

\begin{quote}
\textit{``As parents, we become minorities as well. In my social circle, I'm the only one dealing with this, so I feel like a minority myself'' (P4).} 
\end{quote}

This isolation triggered deep emotional turmoil, marked by fear, confusion, and helplessness, much resembling trauma response as articulated by \citeauthor{chen2022trauma} and \citeauthor{randazzo2023if} \cite{chen2022trauma,randazzo2023if}.

The emotional pain deepened as parents grappled with the realization of what their child must have endured in silence. One parent described, 

\begin{quote}
\textit{``I discovered this enormous secret, but I had no one to talk to about it. I felt so trapped that I thought I would go mad. I had no one to confide in—neither at work, nor among friends, nor even at home. Then, I realized that this must have been the same emotional torment my child felt all these years'' (P10).}
\end{quote}

For some, realizing their child's identity became a reason to begin \textit{listening}, as the abstract concept of queerness suddenly became deeply personal. Under such light, these initial reactions can be seen as preparation and a catalyst for \textit{listening}. By \textit{``becoming a minority together,''} parents began to empathize with their child's struggles, responding to their feelings, needs, and vulnerabilities. At the same time, this shift placed parents in a vulnerable position, challenging their authority and worldviews. This transformation conditioned them to immerse themselves in queer experiences, fostering a newfound openness to understanding their child's reality.

\subsubsection{Seeking Information: Initiation of Recovery}
With little prior knowledge of queer realities, parents grappled with fundamental questions: What does it mean to be gay or transgender? What is this ``queer'' world that my child is part of? Will my child be okay? What do I have to do now? This lack of understanding, combined with concerns and uncertainties about their child's future, motivated them to seek more information about queer realities.

One of the most prominent concerns was health issues, such as sexually transmitted diseases (STDs) often associated with gay men and the potential side effects of gender-affirming therapies that transgender or non-binary individuals may undergo as part of their transition. One parent expressed, 

\begin{quote}
\textit{``There were so many conflicting opinions about hormone treatments. Some said it could have fatal consequences or cause serious side effects. As a parent, I was worried about everything. Should we proceed with this or not?'' (P8).} 
\end{quote}

Not knowing how to move forward amid conflicting health information, along with the fear of making the wrong decision, weighed heavily on parents, prompting them to seek more information on queer-related health information.

Above all, parents were deeply concerned about whether their children could lead happy lives in a society fraught with hatred and discrimination toward queer individuals. As one parent shared: \textit{``I wanted to ask other parents, are our children doing okay?'' (P3).} This concern was compounded by internal struggles, as parents recognized their own biases and the need to educate themselves about queer identities. This internal conflict catalyzed their search for information to reconcile their beliefs and better understand their children.

These concerns and uncertainties motivated parents to engage in active information seeking, which can also be understood as the initiation of their recovery from emotional trauma. After experiencing the emotional toll of learning their children's identity, parents were left in an emotional state of fear and uncertainty; seeking information became a way to make sense of the situation and to find a path forward. Such moments can also be seen as laying the groundwork for \textit{listening}. As \citeauthor{gilligan2014moral} emphasizes, \textit{listening} begins with active inquiry and genuine curiosity about others' stories and concerns~\cite{gilligan2014moral}. Driven by their desire to address their fears and answer their questions, parents were now at a state where they were ready to attentively \textit{listen} to queer realities, contrasting from their initial state of indifference before learning their children's identity.

Taken together, these early moments after learning their children's identities mark an emotional rupture for parents. This places them in a position where they are motivated to \textit{listen} to queer realities in hopes of alleviating fear and uncertainty, prompting them to seek information to understand what queerness is and how to move forward.

\subsection{Seeking and Encountering Queer (Mis)Information: Trauma Response and Coping}

\subsubsection{Lack of Information}

As they began their information seeking, parents often turned to the Internet or books, only to find fragmented or insufficient information. One parent shared, 

\begin{quote}
\textit{“I searched bookstores, flipping through psychology and philosophy sections. The information was vague, often ending in very general and abstract statements that aren't hostile nor helpful like “we cannot tell for sure.” It was unsettling and distressing, and it even made me angry.” (P3).} 
\end{quote}

Another parent echoed this frustration with online searches: 

\begin{quote}
\textit{“I tried various search terms, but the information was neither helpful nor abundant. I couldn't even find someone to talk to, which made it harder. There was plenty of content, but I had no way of verifying its accuracy.” (P8).}
\end{quote}

Parents also shared that reaching out to their children for accurate or lived information was a difficult option. Conversations with their children on this topic carried a heavy emotional toll, and many parents were still unsure whether their children could serve as reliable and unbiased sources of information when they themselves were the subject of concern. As P8 articulated, \textit{“You think that your child might be mistaken, especially if they are still a minor.'' (P8).}

These accounts reflect what \citeauthor{jonas2024better} describe as an information justice problem \cite{jonas2024better}. Social structures, such as Korea's hostile political climate, broader social atmosphere, media landscape toward LGBTQ+ individuals, and indifferent or uninformed experts like teachers, counselors, or psychiatrists, produce \textit{information deserts} for parents. These conditions restrict parents' access to affirming and accurate information about LGBTQ+ lives, limiting their opportunities to meaningfully recover from the emotional toll of fear and uncertainty they experience upon learning their children's identities.

\subsubsection{Misinformation as Graspable, Yet Ineffective, Solutions to Trauma}

In such information deserts, misinformation often became compelling for parents, offering graspable “resolutions” to their uncertainty and fear. A prominent example was parents' belief in misinformation portraying queerness as “curable,” often encountered through materials on conversion therapies\footnote{Conversion therapy refers to practices that attempt to change or suppress a person's sexual orientation or gender identity, often through counseling, religious interventions, psychoanalysis, or coercive techniques such as physical abuse, chemical or surgical castration, or corrective rape. These practices are considered unethical because, aside from their evident cruelty, they are ineffective and have been shown to cause significant psychological harm, including increased risks of depression, anxiety, and suicide, while denying LGBTQ+ people's identities and dignity \cite{bothe2020s}.}. This narrative offered an appealing “way out” of the traumatic situation parents suddenly found themselves in. Health-related misinformation, such as claims that male-to-male sex “creates” HIV or that gender-affirming treatments carry extreme risks, similarly offered what felt like an actionable direction amidst their disorientedness: to pull children away from their queer identities in the name of protection.

These information deserts also meant that many parents did not realize that much of the available queer-related information in Korea needed to be approached with skepticism. In other words, they were often unaware of the presence of queer-related misinformation. Many parents were unaware of the organized groups in Korea actively persecuting LGBTQ+ people, making them less likely to question the accuracy or intent behind the information they encountered. With no institutional education or accurate media representation of queer realities, such misinformation often became their first point of exposure to the concept of queerness. As one parent noted, 

\begin{quote}
\textit{“There's so much information out there, and unless you know you have to check if it's wrong, it's easy to just accept it.” (P7)}
\end{quote}

Although misinformation offered parents graspable and seemingly actionable paths forward, it ultimately deepened rather than alleviated their emotional distress. Much of what they encountered exploited and intensified fears regarding their children's safety. Health-related misinformation, in particular, amplified anxieties. As one parent shared, \textit{“The idea that my child might face diseases like AIDS made me desperate to find a way to protect them.” (P5)} 

Beyond health-related fears, some misinformation suggested that queer individuals inevitably lead unhappy lives, citing suicide statistics or depicting same-sex relationships as purely physical and devoid of love or care—further compounding parents' fear.

Most significantly, misinformation strained parent–child relationships and often generated mutual feelings of betrayal. Some parents came to believe their child had “chosen'' to be queer, grieving imagined futures such as heterosexual marriage or grandchildren—sentiments heightened by Korea's strong filial expectations. Others, misled by misinformation about conversion therapy, pressured their children into unwanted treatments. Though rooted in care, these actions severely damaged trust and communication within families, effectively worsening their emotional rupture.

To sum up, in the absence of reliable sources of information about queer realities, misinformation became an easy option for parents seeking to cope with the emotional rupture of learning their children's identities. Yet, rather than facilitating genuine recovery, misinformation ultimately intensified their fear, confusion, and emotional distress.

This dynamic can also be understood as parents passively \textit{hearing} about queer realities rather than \textit{listening}. In these moments, they turned to the information most accessible to them to manage their own fear, leaving their core worries about their children's well-being unaddressed. Because misinformation offered no hopeful vision of their children's future, however, it ultimately prolonged their distress rather than easing it. This highlights why healing required a relational shift toward \textit{listening}, grounded in genuinely engaging with the realities their children sought to have understood. The following section illustrates how this transformation unfolded.

\subsection{Recovery Through Relational \textit{Listening}: Identity Re-Construction as Supportive Parents}

In the sections that follow, we illustrate how parents \textit{listening} to queer realities through interacting with LGBTQ+ individuals, their families, and narrative-driven media alleviated the fear, isolation, and uncertainty parents experienced. We also show that this recovery was not a simple restoration to a previous state or an erasure of the journey they have taken, but a relational \textit{repair}, ultimately enabling parents to reconstruct their identities as supporters of their LGBTQ+ children~\cite{randazzo2025kintsugi}.

\subsubsection{Interacting with Other LGBTQ+ People and Their Parents.} 

Amid the unresolved emotional toll fueled by information deserts and pervasive misinformation, some parents eventually turned to a more reliable source of information: LGBTQ+ individuals and their families. As one parent shared, \textit{“I tried to find information here and there, but nothing seemed reliable, so I decided I needed to meet other people in the same situation, not just rely on what I was reading'' (P3).} These meetings often took place through PFLAG meetings or connections with their children's partners and friends. 

One parent likened LGBTQ+ individuals to dragons—mythical beings feared and misunderstood because they are rarely seen up close. 

\begin{quote}
    \textit{``Queer people are like dragons. They are distant, and no one has ever seen them. Because no one ever really sees them in real life, they can be easily criticized and feared based on rumors and misinformation'' (P10)}. 
\end{quote}

However, meeting LGBTQ+ individuals transformed these abstract figures into relatable, living people. One parent described how meeting members of a gay rights organization alleviated their fears: 

\begin{quote}
    \textit{``That day, my heart melted. The people I met were so kind, just like my son. By the end of the meeting, and even during the gathering afterward, I felt my anxiety and isolation start to dissolve. I realized that meeting people face-to-face is the best way to gain information. Books are important, but nothing beats human connection. I was healed there'' (P3)}.
\end{quote}
 
 By engaging with queer individuals—asking questions, learning about their experiences, and empathizing with their stories—parents began to ``reveal a common humanity'' between themselves and LGBTQ+ individuals, and by extension, their LGBTQ+ children; an illustrative example of what \citeauthor{gilligan2014moral} identifies as the core experience of \textit{listening}~\cite{gilligan2014moral}.

Additionally, connecting with people who had already navigated similar challenges provided parents with practical, firsthand knowledge that resolved their disorientation. This was particularly valuable for parents of transgender children, who received accurate medical and administrative information from those who had undergone transitions. As one parent shared, 
\begin{quote}
\textit{``Through communicating with people who had experienced it themselves, I learned about the importance of hormone therapy and the processes involved. If we had had this information earlier, my child wouldn't have struggled as much (P8).''}
\end{quote}

Meeting LGBTQ+ individuals also helped parents see that their children could lead happy, fulfilling lives, which provided immense reassurance. This realization dispelled much of the fear instilled by misinformation, as parents began to envision hopeful futures for their own children. As one parent explained, \textit{``Meeting other queers made me see how well they are doing, living well. So, that's where I found hope (P3).''}

In summary, direct in-person interactions with LGBTQ+ individuals and their families offered insights grounded in lived experiences, which proved invaluable for alleviating their emotional distress rooted in fear of the child's health or future. These encounters allowed active \textit{listening} to take place, dismantling prior misunderstandings and revealing the humanity in those who had once been seen as ``dragons.''

\subsubsection{Narrative-based Media as a Source for \textit{Listening}.} 

Novels, movies, and TV shows served as another powerful avenue for \textit{listening}. These narratives often depict the profound emotional journeys of discovering one's identity, finding love, and navigating challenges such as coming out to family members. Media representations that move beyond superficial or stereotypical portrayals and authentically engage with queer lives allowed parents to emotionally connect with the experiences of queer individuals.

One parent reflected on how reading the novel \textit{Make Me One Dimensional} by Park Sang-young, which explores a gay character's life, brought back memories of their own adolescence, dispelling the misinformation that same-sex relationships are only about physical lust~\cite{park2021want}. She shared, 

\begin{quote}
\textit{``People often say that homosexuality is just about physical relationships or fleeting pleasures, but as I read that novel, I felt a deep tenderness. It made me realize that queer love is no different from heterosexual love. It felt precious to me'' (P2).}
\end{quote}

These narratives offered a unique avenue for \textit{listening}. Though not direct human-to-human interactions, engaging with these stories helped parents see LGBTQ+ individuals as real people rather than abstract subjects. By bringing queer experiences to life, the narratives fostered empathy and deeper understanding. They enabled parents to grasp the significance of sexual identity and orientation for their children, challenge common misconceptions—such as the idea that same-sex relationships are purely physical—and develop a more nuanced perspective on queer lives.

\subsubsection{\textit{Listening} to Children} 

Another vital source that parents \textit{listened} to was their own children. This was not necessarily a direct conversation with their children, but rather a realization and deep resonance with who their children truly were as queer individuals and what that meant for them. P4, for instance, described the moment her child awoke from breast surgery with immense joy—a moment that reshaped her understanding: 

\begin{quote}
\textit{``After the breast surgery, when the gauze was removed, I… I couldn't look at it. So I looked at my child's face instead, and it was… Sheer joy itself. It was at that moment that I knew my fears were misplaced. Face filled with joy, despite having those 10cm wounds on the breasts, saying they felt no pain... I realized it was my own greed to feel something about it at all. I completely accepted after that'' (P4)}.
\end{quote}

These moments, grounded in pre-existing bonds of parental care, helped parents empathize and see the world through their children's eyes. Their love fostered a kind of \textit{listening} that immersed them in queer realities and deepened their understanding of their children's lives. This empathy became a foundation for recovering from their fear of whether their children would be happy as LGBTQ+ individuals. They came to understand that being queer and being recognized as their genuine selves was at the center of their children's happiness. As one parent realized, \textit{“this is not a matter of right or wrong, but something that just needs to be accepted.” (P10).}

\subsection{Transformed Informating Practices after Identity Reconstruction}

In the previous section, we showed how parents began recovering from emotional rupture through relational and emotional \textit{listening} to queer realities, a process that reconstructed their identities as supportive parents. In this section, we examine how this renewed identity reshaped their information-seeking and validation practices, most clearly seen in how they perceived, guarded against, and actively challenged the misinformation that had previously fueled their anxiety and fear.

\subsubsection{Critically Analyzing (Mis)Information}
As parents began resonating with queer individuals, they came to recognize that the misinformation they encountered may not have been driven by genuine concern toward LGBTQ+ individuals (e.g., ``saving'' queer people from STDs or unhappy lives). This fostered a meta-awareness and critical analysis of the (mis)information they are presented with. For instance, when presented with an example misinformation article that we prepared, this is how P7 responded:

\begin{quote}
    \textit{“There's that thing called ‘fear-mongering.' Here, it actually seems like the intention to instill fear is even more deliberate. When I look at these articles, things like alcohol dependence or suicide risk... Yes, there is some truth in those aspects, right? But when it comes to the hardships of life itself, and the constant invalidation of one's very existence as a minority… When someone receives that kind of constant feedback, it becomes extremely difficult to cope. So, of course, the risk of suicide or similar issues would be higher. That's how I see it. But if that's the case, the real cause of such problems lies elsewhere, rather than being queer itself. And yet these articles don't mention anything about those causes. So I end up thinking, for society to support people in these situations, certain things need to change, certain prejudices… But without any mention of that, the article just feels like it was written purely to instill fear.”} (P7)
\end{quote}

This realization encouraged a more critical stance toward the information they encountered. Parents reported how they started noticing the exaggerated language used to distort queer realities when reading queer-related articles. For instance, P7 observed, \textit{``I felt that they were emphasizing certain aspects to make them seem worse than they are.''} Others questioned claims of curing homosexuality, asking how such stories were verified: \textit{``Were these people really gay? If so, what proof exists that they were `cured'? '' (P9)}

\subsubsection{Protecting Oneself from Exposure to Misinformation}
Further, when encountering queer-related content on news portals or social media, many participants reported developing an ``antenna'' to detect harmful narratives. Aggressive headlines, unsupportive tones, or media sources known for homophobia were often enough for parents to instinctively dismiss the content and stop themselves from even reading or deeply engaging with these articles, protecting themselves from emotional damage. P3 perfectly showcased this case when we presented them with a sample misinformation article for exercise: 
\begin{quote}
\textit{``You want me to discuss the content of this article? I don't need to read it. I just don't read this stuff. Look at the title!'' (P3)} 
\end{quote}

P5 also shared that during the Mpox outbreak, when Korean media blamed gay men for the spread of the disease, she initially felt concerned but quickly dismissed these claims based on her prior experiences with HIV/AIDS-related rhetoric targeting gay men.

Some parents also creatively leveraged algorithmic environments to their advantage. P2, for example, engaged with queer-affirming YouTube content to shape her recommendation feed, minimizing exposure to harmful material: \textit{``There is a limit to finding this information alone. But this, it just does it for me'' (P2)}. While algorithmic bubbles are often critiqued for reinforcing bias~\cite{bryant2020youtube}, this case illustrates how they can also be harnessed to create safer, more supportive information spaces.

\subsubsection{Becoming an Agent in Queer Information Space: Creating Community of Care and Fighting Against Misinformation}

Having witnessed its harm firsthand, some parents felt compelled to actively counter misinformation. As one parent reflected, \textit{``If we'd had the right information, my child wouldn't have suffered alone'' (P8)}. This realization motivated some of them to advocate for queer realities and confront harmful narratives. These actions can be understood as novel informational activist practices rooted in their newly reconstructed identities \cite{jonas2024better}.

One such practice was establishing a safe information community. Many who had benefited from the PFLAG community felt a strong responsibility to pay this support forward, sharing YouTube videos, books, and articles with newcomers and family members. As P4 explained, \textit{``If there are young LGBTQ+ individuals out there, I want to share what I've learned and help other parents avoid the mistakes I made'' (P4)}. 

Some parents even extended this practice to our research team, recommending resources during and after interviews. Through such engagements, parents nurtured bonds among PFLAG members and within their families, cultivating an “informational care community'' that embodied solidarity across online and offline spaces.

Parents also took active steps to confront misinformation directly, both individually and publicly. P10 shared, \textit{``I have conversations where I share that my child is queer and that I'm involved in these activities (…) It helps people see that queer individuals are just neighbors and family members like anyone else.''} 

Others amplified queer voices through media: \textit{``(In PFLAG,) We work hard to create YouTube content, write books, and give interviews to counter misinformation and ensure that people aren't misled'' (P3)}. Some even organized collective actions against misinformation circulated by media outlets: \textit{``We decided to protest as a group because it's more effective than individual complaints… We called the broadcasting system to tell them they were wrong'' (P1).}

Taken together, our participants illustrate how individuals who initially experienced the emotional rupture of learning their children's queer identities—and were therefore vulnerable to misinformation—gradually reconstructed their identities through a relational healing journey. Through this process, they shifted from overwhelmed recipients of (mis)information to critical and protective information consumers, and in some cases, active advocates who counter harmful narratives. In other words, their roles, identities, and informating practices have shifted in the informational infrastructures surrounding queer realities in Korea \cite{yang2019seekers}. With these newly developed informational practices, parents who were once \textit{listeners} now serve as powerful storytellers for both other parents in similar situations and LGBTQ+ individuals navigating strained family relationships.

\section{Discussion}
Our findings reveal how parents' fears and uncertainties upon learning their children's identities gradually healed through the process of \textit{listening} to queer realities, enabling them to reconstruct their identities as supportive parents and, in turn, transform their informating practices. These include becoming more critical in assessing queer-related (mis)information, cultivating strategies to protect themselves from harmful narratives, and actively challenging misinformation to support others navigating similar experiences. 

In addition, our case illustrated that misinformation navigation constitutes a complex sociotechnical journey, far beyond the scope of individual fact-checking, resonating deeply with prior work on the relational and infrastructural dynamics of misinformation \cite{juneja2022human, allen2022birds, varanasi2022accost, hassoun2023practicing, huang2015connected, gao2018to, scott2023exploring, vraga2017using, vraga2020correction}.

\textcolor{blue}{In our discussions, we first critically reflect on the limitations of our study.} We then examine the relational and informational conditions that enabled \textit{listening} to take place, highlighting the importance of informational presentations that “humanize'' queer realities rather than abstracting them. We further propose several future research directions and, finally, discuss our positionality as researchers within a broader research landscape.

\subsection{Critical Reflection on the Findings}

Before diving into broader discussions, we begin with a critical and cautious reflection on the study's limitations, particularly its limited scope. Our findings draw from \textit{supportive} parents of LGBTQ+ individuals. This recruitment choice was intentional, as our goal was to understand the parental recovery journey. Yet, this also means our findings trace a hopeful trajectory that may not reflect the difficult realities many LGBTQ+ individuals face, while parental support remains unattainable for many~\cite{haas2010suicide, king2008systematic, richter2017examining, grossman2021parental}. 

That said, in presenting our cases, we do not intend to overlook or minimize the pain experienced by those with unsupportive families, nor to generalize our findings beyond the situated contexts we examine. Instead, we offer this work as an invitation to center parent-child relational dynamics as a critical dimension of queer-centered research, and we encourage the CSCW community to further explore this important yet under-examined aspect of queer well-being.

Additionally, while we highlight the importance of empathic resonance with queer lived experiences in countering misinformation, we acknowledge the risk of suggesting that lived experience is the sole or definitive basis of truth. Centering queer voices is essential, especially in contexts where they are marginalized or erased, but lived experience is neither infallible nor universally representative. For instance, in medical or clinical settings, trained professionals may offer critical insights that complement or challenge individual narratives. While we remain firmly committed to affirming the value of LGBTQ+ lived experience, we advocate for balanced approaches that integrate multiple forms of knowledge. 

\textcolor{blue}{Finally, we critically reflect on how \textit{listening} was used in our paper. \textit{listening} was not used as an a priori analytical framework during data collection. Participants were not directly asked about their \textit{listening} practices or their role throughout their journeys. Instead, \textit{listening} emerged as a recurring pattern in participants' accounts, which we later used to interpret and present our findings. As such, while we argue that \textit{listening} provides a useful lens for understanding the parental recovery journey, as well as identity and informating practice transformation, we do not claim a direct causal relationship. That said, we look forward to future work that further examines misinformation concerning marginalized populations through a care ethics lens and \textit{listening} as a theoretical grounding.}

\subsection{Conditions That Enable \textit{Listening}: Reflecting on Modes of Information}

\subsubsection{Humanizing Encounters vs. Abstract Encounters}
Having shown how parents healed from emotional rupture through relational \textit{listening}, we now turn to a deeper conceptual question: what condition made such \textit{listening} possible? In this section, we unpack this question by foregrounding the role of humanizing, courageous, and relational encounters between subjects and consumers of information (i.e., LGBTQ+ individuals and parents with LGBTQ+ children, in our context). We contrast these encounters with abstract, decontextualized forms of information that fuel misinformation and deepen trauma.

We see the coming together of parents and their LGBTQ+ children as a series of \textit{encounters}, as defined by \citeauthor{hoseason2010role}, in which oppositions and distinctions are enacted, asserted, and negotiated, with both positive and negative possibilities \cite{hoseason2010role}. In the context of our work, we observed several types of encounters between parents and queer realities: 

\begin{enumerate}
    \item \textbf{In-Person Meetings}: Face-to-face interactions where parents meet other LGBTQ+ individuals, such as through PFLAG meetings or meetings with their children's partners and friends.
    \item \textbf{Conversations with Children}: Direct dialogues in which children openly share their identities and personal journeys with their parents.
    \item \textbf{Exposure to Information that Reflects Lived Reality}: Engagement with sources that reflect the lived realities of queer individuals, including narrative-driven media such as films and literature, as well as personal stories shared through blogs or YouTube vlogs by LGBTQ+ individuals.
    \item \textbf{Exposure to Descriptive Information About Queerness}: Engagement with sources that present queerness in a more detached or informational manner, such as academic papers, news articles, or explanatory texts.
\end{enumerate}

Of these encounters, our findings show that the first three were where active \textit{listening} often took place for parents, while misinformation was predominantly presented in descriptive form. What distinguished these successful encounters from those with descriptive information are two critical factors: (1) \textit{humanization}, where queer people were brought out from a distant and abstract representation to living, breathing individuals in front of them, and (2) \textit{courage}, where both parties in the encounter actively placed themselves in front of each other, faced each other's vulnerabilities and connected \cite{gilligan2014moral}.

In terms of humanization, care ethics emphasize the power of \textit{listening} in manifesting the common humanity among participants in a relationship \cite{gilligan2014moral}. Our findings emphasize how parents engaging with LGBTQ+ people in person is a completely different experience from trying to understand what queerness is based on descriptive information. Not only can one directly ask questions and hold conversations, but one also sees the other as a living, breathing individual. The actions that go beyond textual communications---A child crying during a coming-out, the joy on a child's face after gender-affirming surgery, an actor subtly expressing their vulnerability in queer films through their facial expressions, or the shared laughter and tears at PFLAG meetings---become a powerful avenue of communication where queer individuals transform from abstract subjects or statistics into real people. This humanization sharply contrasts with the portrayal of queer people in descriptive misinformation, where they are often reduced to scandalous anecdotes, dehumanizing images, or impersonal statistics.

Courage also plays a crucial role in these encounters. Both parties—parents and children—must be vulnerable and open for empathic \textit{listening} to take place. Children bravely share their identities and personal journeys, while parents confront their internal prejudices and the challenging reality of accepting their child's queer identity. These decisions are not to be taken lightly; both parties actively choose to step outside their comfort zones with a deep desire to understand and be understood by one another. In contrast, in many examples of misinformation, such courage is absent. These sources were often crafted with the intent of imposing their own views on the public, rather than having the courage to face discomfort and engage with the unknown.

The type of information we are engaging with here is a very specific kind: information that attempts to describe a way of being by those who are not of the same identity. We observe that such descriptive forms of information are more susceptible to hostile misinformation, particularly when they are disembodied from the people being described and abstracted into distant beings, offering limited grounds for genuine \textit{listening} to take place. In contrast, other ways of encountering—those grounded in human interaction, empathy, and \textit{listening}—can enable parents to see queer individuals as fully human, just like themselves.

\subsubsection{Envisioning Humanizing Interactions}
What does this insight offer CSCW scholars and practitioners, whose work often centers on connecting people across different ways of being through technological systems? A key question we must ask is this: what forms of information do current social technologies amplify? We argue that much of online informational infrastructure today still relies heavily on textual forms—posts, comments, messages, threads—that can abstract people's experiences into distant, depersonalized fragments. In such formats, it becomes increasingly difficult to recognize the person on the other side of the screen, or to treat people described as living, breathing human beings like the consumers of information themselves.

In contrast, communication rooted in humanization, where the “other” is encountered not merely as a textual representation, but as someone who breathes, laughs, and lives, offers a more powerful foundation for mutual understanding. To design online spaces that allow for such humanization, we might envision embracing communicative modes that foreground embodied forms of information transfer to surface shared humanity among participants. This also resonates with \citeauthor{weiser1991computer}'s notion of \textit{embodied virtuality}, where he envisioned that “by pushing computers into the background, embodied virtuality will make individuals more aware of the people on the other ends of their computer links” \cite{weiser1991computer}. In other words, it is about helping people realize that those on the other side of the screen are \textit{humans} like themselves.

An illustrative case of such a humanizing encounter is Clubhouse, a voice-based social platform that gained popularity around 2021~\cite{clubhouse}. Clubhouse enabled users to join real-time conversations only via voice, fostering interactions that felt more embodied and emotionally rich. Remarkably, even amid polarizing political contexts, such as the Palestine–Israel conflict, Clubhouse facilitated conversations in which individuals from both sides came together to speak, listen, and connect in meaningful ways~\cite {adams2021clubhouse}. 

We speculate that, for speakers, the embodied experience of \textit{saying} something out loud may introduce more self-awareness than the disembodied experience of \textit{typing}. For instance, a “fxxk you” text may be easier to type and send than saying it aloud in voice, where the affective weight is felt more viscerally. For listeners, hearing a person's voice—complete with pauses, breaths, and tonal nuance—can feel more personal than seeing the same thing written on a Reddit forum. 

A compelling parallel can be drawn to \citeauthor{gilligan2014moral}'s discussion of the documentary \textit{The Gatekeepers}~\cite{moreh2012gatekeepers}, where a former Israeli intelligence official reflects on their interactions with those they once perceived as enemies. After a direct encounter, one official remarked: “I see you don't eat glass. He sees I don't drink petrol”. This illustrates how direct encounter enables mutual recognition of their own humanity and gives rise to care~\cite{gilligan2014moral}.

Another way to approach this insight is to recognize the possibly inherent limitations of today's informational technology landscape. Much of our information infrastructure has been shaped by workplace-oriented values of efficiency and productivity \cite{suchman1987plans}. As a result, the dominant platforms and media environments may have evolved in ways that privilege abstract, decontextualized representations of information, enabling fast-paced information exchange. It is therefore difficult (if not structurally constrained) to expect these technologies to facilitate the kinds of humanizing, embodied encounters naturally. 

In such cases, the more meaningful question may not be on how technology itself can humanize, but where the constraints lie and how it may aid conditions or infrastructures that make humanization possible—such as facilitating in-person gatherings, supporting community-organized encounters, or enabling access to trusted human intermediaries—rather than assuming that technology must serve as the primary medium of humanizing engagement.

To conclude, we reflect on how abstract informational representations may constrain our ability to recognize one another as fully human. We invite CSCW scholars to explore humanizing modes of communication as pathways toward more inclusive, caring, and emotionally resonant sociotechnical environments.

\subsection{Exploring Design Possibilities}

In this section, we offer \textit{design possibilities} for supporting parents of LGBTQ+ individuals as generative directions for future inquiry. We frame these not as prescriptive \textit{design implications}, but as exploratory \textit{possibilities}, mindful that they are early design ideations inspired by our data rather than fully developed concepts grounded in extensive design research. We stay cautious against proposing premature or overly concrete systems that might risk overlooking the multifaceted contexts that shape these experiences, including intergenerational trauma and local queer landscapes \cite{dourish2006implications}. We also aim to avoid communicating a techno-solutionist vision in which a particular system is presented as capable of “solving” the challenges at hand, especially when much of our paper has shown how these journeys involve complex interminglings of societal, interpersonal, and non-systemic factors.

In presenting our design possibilities, we adhere to the idea of augmenting the already well-executed practices that our participants showcased. This aligns with \citeauthor{wong2020culture}'s use of Culture in Action theory by \citeauthor{swidler1986culture} \cite{wong2020culture,swidler1986culture}, as well as the notion of celebratory technologies proposed by \citeauthor{grimes2008celebratory} \cite{grimes2008celebratory}. These approaches highlight how designs may benefit more from focusing on people's existing capacities and assets rather than their needs and deficits, maximizing their lasting impact by accounting for how people already live rather than narrowly targeting a specific problem. Informed by these perspectives, we revisit the specific strategies parents currently use to navigate their journeys and propose design possibilities that may augment those practices as meaningful directions for future work.

\subsubsection{Enabling Agency as Informational Actors}

Our participants exercised agency as informational actors in multiple ways, such as developing “antennas'' to detect content likely to trigger emotional trauma or strategically using YouTube's recommendation system to both avoid re-traumatizing materials and surface helpful or queer-affirming information. These tactics reflect a form of “make-doing'' within their informational environments, leveraging their own abilities alongside platform affordances.

Building on this observation, we advocate for systems that further support and expand such forms of agency and subjectivity \cite{bardzell2015user}, particularly by reducing the risk of re-traumatization when encountering hostile or stigmatizing content \cite{chen2022trauma,randazzo2023if}. One possibility is the addition of more explicit guided-encounter features to informational platforms. Rather than relying on opaque recommendation feeds, platforms could provide transparent, structured pathways that allow parents to curate what they wish to see more or less of in media portals where algorithmic recommendations shape much of their informational experience.

\subsubsection{Community-Driven Informating Practices}

Our findings highlight the importance of community-driven practices during their recovery journey. Parents described how PFLAG offered practical guidance (e.g., where to find queer-friendly hospitals or credible resources), emotional validation that reduced feelings of isolation, and humanizing encounters that helped them confront internal discomfort with queer identities, echoing the dynamics of information activism described by \citeauthor{jonas2024better} \cite{jonas2024better}.

While PFLAG Korea is a central space for such sensemaking, it also faces structural constraints. Drawing on two participants who were not involved in PFLAG, alongside the first author's personal conversations with other parents outside of this study, we note several barriers: the main office is in Seoul and regional chapters remain limited, making access difficult for those outside the metropolitan area; some parents worry that attending meetings may expose them as “parents of queer children,” risking social scrutiny in a highly stigmatized context \cite{laurent2005sexuality}; others hesitate because they do not feel ready to engage in advocacy; and in-person attendance itself can require significant emotional readiness for parents still navigating early stages of acceptance.

These realities reveal a gap between the need for community-driven sensemaking and parents' uneven ability or willingness to engage with existing support groups that require physical presence and identity disclosure. In these moments, technologies may offer “soft'' forms of complementary support. Potential designs include online communities that enable anonymous collective sensemaking between new and veteran parents, video-based support groups with privacy-protective features such as AR filters \cite{noh2024investigating,noh2025counselar}, or collaboratively maintained databases of queer-friendly hospitals, counseling centers, books, and media. Such systems could provide emotionally and geographically lower-barrier entry points into supportive networks for parents who feel uncertain, distant, or reluctant to join formal advocacy spaces.

\subsubsection{Centering Parents as Those in Need of Help rather than Sole Bearers of Caregiving}
Our findings also highlight the importance of recognizing parents' own journeys as meaningful, rather than treating them as peripheral to LGBTQ+ individuals—a theme echoed in both PFLAG communities and participants' reflections. Many parents expressed that, in hindsight, they know they “should” have been supportive from the beginning. At the same time, they reflected on how their responses were shaped by their own upbringing and by cultural norms in Korean society, which ultimately subjected them to a significant emotional toll when confronted with their children's self-disclosure as queer, as well as feelings of regret over actions they took toward their children due to a lack of understanding.

Of course, we do not raise these accounts to excuse or legitimize abusive or toxic parental dynamics. Rather, we highlight them to underscore the importance of \textit{listening} to parents' struggles from their situated perspectives—not only to support meaningful change, but also to recognize them as individuals who experience their own hardships and emotions as parents of queer children. Centering parents, therefore, should move beyond the instrumental logic of “helping them understand LGBTQ+ individuals for their children's sake'' and toward acknowledging parents as people who may also need attention, guidance, and care. From this standpoint, design possibilities could more explicitly \textit{center} parents, such as by providing mental health or crisis support tailored to their needs.

\subsubsection{Critical Reflections on Design Possibilities}

Before we conclude this section, we ask a critical question: will any of these ideas truly “solve'' the challenges faced by parents of LGBTQ+ children, or by LGBTQ+ individuals with hostile parents? As discussed in our related work, navigating inter-generational trauma and countering hostile misinformation is profoundly complex. No single platform, technological intervention, or policy change can address the structural and societal conditions shaping these experiences. What happens, for instance, when parents have no desire to engage with such resources at all? Further, many challenges stem from what we called informational deserts in our findings: the absence of queer-inclusive curricula, skewed portrayals of LGBTQ+ individuals in mainstream media, and the lack of legal protections against hate speech and discrimination in Korea. These are issues that require broad social, educational, and political efforts, none of which can be resolved by a simple design intervention.

Thus, it may be necessary to acknowledge that these issues cannot be “solved'' outright, but must instead be approached through ongoing, situated, and multi-level efforts~\cite{harrington2019deconstructing}. We offer these reflections not to undermine the design possibilities discussed above, but to present our findings responsibly by recognizing our limitations. Even so, we believe that continued advocacy across diverse sectors is what ultimately fuels change. As we present our analysis of the sociotechnical entanglements surrounding queer realities, we hope that the value of this study extends beyond outlining future design possibilities and contributes to inspiration, momentum, and hope for our community in shaping a more caring world.

\subsection{\textit{Listening} as a Researcher: Reflecting on the Researcher-Participant Relationship}
To conclude our study, we shift focus to reflect on our role and experiences as researchers conducting this work. Throughout the research process, we came to recognize that our endeavor was itself an act of \textit{listening} to the lived realities of parents of LGBTQ+ individuals in Korea. \citeauthor{toombs2017empathy} emphasize that in studies where researchers and participants deeply engage with one another, a genuine caring relationship forms \cite{toombs2017empathy}. This relationship not only shapes the framing and outcomes of the research but also fosters emotional solidarity that extends beyond the study itself. This mutual empathy, as previously discussed, emerges when both researchers and participants show vulnerability and connect through a shared sense of humanity~\cite{howard2019ways}.

Reflecting on our research process, we found that such vulnerability was indeed prevalent throughout our study. Not only did parents open up about their stories, but the first author, who conducted the interview and is a gay man around the same age as many of the participants' children, often found himself tearing up together with the participants while listening to the parents' stories and sharing his own experiences to help their children. Following moments like these, where both researcher and participants opened up to each other, the first author noticed himself receiving invitations to PFLAG events, book and movie recommendations, and warm expressions of gratitude, with participants saying, ``Thank you for trying to make our voices heard.'' In a way, parents actively invited the researchers into their community of care. As \citeauthor{toombs2017empathy} articulates, ``the relationships extend(ed) beyond a mutual interest in the research project'' \cite{toombs2017empathy}.

As an advocacy-oriented research effort aimed at amplifying marginalized voices~\cite{creswell2017research}, we underscore the importance of nurturing such caring relationships beyond the scope of the research itself. Through conducting research that advocates for marginalized realities, researchers inevitably become part of the participants' care community, joining them in creating spaces where their voices are genuinely \textit{listened} to and valued through our work. Aligning closely with community-based participatory and action research approaches~\cite{cooper2022systematic, le2015strangers, lu2023participatory, hayes2011relationship,harrington2019deconstructing}, we highlight the critical need to go beyond mere observation or reporting. Researchers must deeply engage with marginalized voices, empathize with their experiences, and take responsibility for fostering meaningful connections that extend beyond the research results. Only when this caring relationship between researchers and marginalized communities forms a united front against systemic oppression can we claim to have conducted deeply sincere research, with a real and lasting impact on the world.

\section{Conclusion}
In this paper, we traced the journeys of Korean parents with LGBTQ+ children as they navigated the emotional rupture that followed their children's self-disclosure, their encounters with (mis)information, and their subsequent identity reconstruction as supportive parents grounded in relational \textit{listening} to Korean queer realities—an orientation that reshaped their informating practices. Taken together, our work positions parents as critical yet understudied stakeholders within CSCW.

\begin{acks}
We are deeply grateful to the parents who participated in this study and to the members of PFLAG Korea who contributed in making this study possible. Beyond this paper, we extend our respect and appreciation for their continued efforts toward creating a more inclusive and compassionate society for LGBTQ+ individuals and their parents. We also thank Catherine Wieczorek, Anh-Ton Tran, Seongjo Jeong, Seora Park, and members of KALQ for their generous guidance and encouragement throughout various stages of the project. We are also grateful to friends JongHyeon Baek and Mangnani for their kind assistance with participant recruitment. Lastly, we extend our heartfelt thanks to the anonymous reviewers for their valuable feedback.
\end{acks}

\bibliographystyle{ACM-Reference-Format}
\bibliography{01reference}

\appendix

\section{Articles Used for Misinformation Activity}
\textit{Please note that the original articles were published in Korean and were presented in Korean to our interview participants. We used ChatGPT to translate the articles and first author revised them to check the correctness of translations.}

\subsection{Homosexuality is a Cause of Diseases - Homosexual Acts Are the Main Cause of AIDS... Helping People Quit Homosexuality is True Love (GBN News) \cite{article1}}

[…] As male homosexuals age, their sphincter muscles weaken, causing them to go to the bathroom more than ten times a day and wear diapers. Male homosexuals who engage in anal intercourse are prone to diseases that are uncommon among the general population. In Korea, 21\% of syphilis patients are homosexuals. […] Hemorrhoids, bleeding, rectal cancer, sexually transmitted diseases, hepatitis, and AIDS are the results of abnormal sexual activity that goes against the structure of the human body, such as anal intercourse.

[…] According to the cumulative statistics on infection routes since 2006, sexual contact accounts for 99.9\% of HIV infections, vertical transmission for 0.07\%, and shared drug injection for 0.03\%. […] The data showing that 93\% of the cumulative total of AIDS patients are male, and that approximately 93\% of new AIDS patients are male, clearly indicates that AIDS is predominantly transmitted through male homosexual activity.

[…] Due to the very high rate of HIV carriers among male homosexuals, 31 countries have completely banned blood donations from male homosexuals. Many European countries are included in this group, while 13 countries, including Korea and the United States, allow blood donations conditionally.

[…] The lives of homosexuals are not as happy as portrayed in movies or dramas. […] 60\% of male homosexual relationships end within a year, and female homosexual relationships end within three years. The average duration of relationships for male and female homosexuals is about 2.5 years, with it being rare for them to last more than five years.

Homosexuals experience about twice as many difficulties related to sexual issues as heterosexuals. Even after AIDS was discovered, and despite receiving education about AIDS and witnessing their friends die from it, homosexuals continue to engage in sexual relationships with strangers, indicating that their sexual behavior is addictive. […] Additionally, homosexuals suffer from loneliness as they age because they often lack family. Four independent studies conducted between 1998 and 2001 found that homosexuals are at least twice as likely as heterosexuals to be dependent on alcohol, and male homosexuals are three times more likely to attempt suicide than male heterosexuals. The risk of cancer among AIDS patients is 20 times higher than that of the general population. Due to AIDS and other diseases, the lifespan of male homosexuals is 25 to 30 years shorter than that of male heterosexuals and 5 to 10 years shorter than that of alcoholics. These findings show that the life of a homosexual is far from happy. Encouraging someone to continue a homosexual lifestyle is not true love. Helping them break away from homosexuality and live a normal life is the true way to support them.

\subsection{Sex Reassignment Surgery Leads to Lifelong Suffering... Suicide Rate is 22 Times Higher (The Mission News) \cite{article2}}

Recently, a study was released indicating that thousands of minors have undergone surgery to establish a transgender identity. Transgender advocates have consistently argued that minors only take puberty blockers or cross-sex hormones and do not undergo surgeries to alter their bodies to resemble those of the opposite sex.

However, the \textit{Journal of the American Medical Association} reported that over the past five years, more than 3,000 minors in the United States have undergone transgender surgeries, with around 400 of them actually undergoing procedures to alter their bodies to resemble those of the opposite sex. It is also an open secret that the number of sex reassignment surgeries among adolescents in our country is continuously increasing.

[…] A considerable number of people suffer from complications following sex reassignment surgery. Even after undergoing such surgery, one's original body's sex, brain, and concepts do not change, leading to a lifetime of regret and mental anguish.

It is said that most individuals who undergo sex reassignment surgery regret it and have a suicide rate 22 times higher than the general population. There are also reports that 40\% of people who attempt suicide in the United States are transgender individuals. Looking at cases from other countries, it is evident that the high suicide rate among transgender individuals is not due to social discrimination but rather the mental and physical aftereffects of sex reassignment surgery.

Johns Hopkins Hospital in the United States, which is at the forefront of sex reassignment surgery, reportedly delayed surgery for minors by providing counseling, leading to 80\% of them abandoning the idea of surgery. The majority of transgender impulses during adolescence are temporary, and there is a high possibility that they will return to normal over time. Therefore, the best way to genuinely support adolescents experiencing transgender impulses is to guide them away from choosing this dreadful surgery.

\subsection{Homosexuality is Not Love. They Grow Old Alone and Eventually End Up Miserable: Testimony of Korea's First Transgender Person Kim Yu-bok on Their Deathbed (Kukmin Daily) \cite{article3}}

Kim Yu-bok (75), a "living witness" of ex-gay life, is struggling between life and death in the ICU at Soonchunhyang University Hospital in Seoul. […] Kim was one of the early transgender individuals in Korea. Even in the 1960s, when the term "transgender" was not used, there were cross-dressing men and women. However, after a gay club for homosexuals was established in Itaewon, Kim came out and became known as the first transgender person to publicly live in Korea. Although he did not undergo sex reassignment surgery, he confessed, "I grew up without receiving love, so I didn't even know the meaning of love and just lived day by day, driven by lust."

From 1960 to 2004, Kim sang at various gay clubs under the name "Kim Marine." For him, singing was his life and everything. However, his life changed drastically in 2004 when he underwent surgery for scoliosis. […] After the surgery, Kim was left unable to walk. When he became disabled, no one visited him. As a basic livelihood recipient, he continued to live a hellish life in a tiny room, barely large enough for one person, surviving on government welfare. Volunteers cared for him, handling even his basic bodily needs.

By his side during this lonely time was Pastor Lee Yona (66, head of Holy Life). Pastor Lee also lived as a homosexual and ran a gay club in Itaewon until he was over 40. Kim performed as a singer at the gay club that Pastor Lee managed. After learning that her son was gay, Pastor Lee's mother took her own life. This tragic event led Pastor Lee to desperately try to escape homosexuality. He eventually went to Japan, studied theology, overcame his homosexuality, and returned to Korea.

In 2015, Kim and Pastor Lee appeared together in a documentary produced by Holy Life (Ex-gay Human Rights Forum) titled "I Am No Longer Gay." The documentary aimed to raise awareness of the pain and suffering experienced by homosexuals and their families and to deliver the message that "homosexuality can be healed." The 1-hour-and-8-minute documentary surpassed 95,000 views on YouTube within nine days of its release. In the documentary, Kim testified to the tragic end of his life. He emphasized, "I feel really sorry for homosexuals. They may experience brief physical pleasure, but that is not love."

He also confessed, "The end of homosexuality is miserable. You can't get married, you grow old and ugly. Friends around me died from AIDS or committed suicide. The end of homosexuality is nothing but loneliness, with no one around. I didn't realize at the time that it was a mistake." […]

\subsection{Case Studies of Homosexuality Treatment (Credo Association) \cite{article4}}

First, let's examine the claims of experts involved in the treatment of homosexuality. Dr. Bieber, after conducting a 20-year study, claimed that the possibility of changing from homosexuality to heterosexuality ranged from approximately 30\% to 50\%. He also revealed that, based on observing his treated patients over a 10-year period, they remained heterosexual. Masters and Johnson reported a 71.6\% success rate six years after treating 67 homosexual men and 14 lesbians. Psychiatrist Dr. Wilson claimed a 55\% success rate when treating homosexuals who were Christians. Clinical psychologist Dr. Kronemeyer stated that approximately 80\% of homosexual men and women were transformed into healthy and satisfied heterosexuals after treatment.

References:
\begin{itemize}
    \item Bieber, I. \textit{Homosexuality: A Psychoanalytic Study} (New York: Basic Books, 1962).
    \item Bieber, I., and T. B. Bieber, "Male Homosexuality," \textit{Canadian Journal of Psychiatry} 24(5), 416, 1979.
    \item Masters, W. H., and V. E. Johnson, \textit{Homosexuality In Perspective} (Boston: Little, Brown and Company, 1979).
\end{itemize}

\subsection{"Homosexuality Treatment Research Shows 79\% Average Effectiveness"... Overturning the "Innate" Claim (Kukmin Daily) \cite{article5}}

[…] Those who advocate for the prohibition of conversion therapy for homosexuality present two main arguments. First, they claim that since homosexuality was removed from the list of diseases (DSM-III) by the American Psychiatric Association in 1973, it should not be considered a disease that requires treatment. However, even if a condition is not classified as a disease, people should still have the right to seek treatment or counseling if they experience discomfort or unhappiness.

They also point out that since 2009, the American Psychological Association and the American Psychiatric Association have argued that conversion therapy for homosexuals is ineffective and actually causes harm to the mental health of homosexual individuals. Instead of conversion therapy, they recommend affirmative psychotherapy, which is intended to provide support and help individuals build self-esteem. However, these claims are based on the incorrect assumption that “homosexuality is innate.” The claim that homosexuality is inherited was ultimately discredited by advanced genetic research conducted on 470,000 people, the results of which were published in the journal \textit{Science} in 2019.

[…] Meanwhile, researchers such as Ramafedi (1992) and Diamond (2003), who are themselves homosexual, have reported that homosexuals often naturally experience changes in their same-sex attraction or homosexual identity as they age, particularly after adolescence, even without efforts to undergo conversion therapy. A study using U.S. population data found that more than 2\% of the population naturally changed their “sexual orientation” over a 10-year period. This phenomenon is referred to as “sexual orientation fluidity.” This counters the claim that homosexuality is innate and therefore never changes over a lifetime.

[…] There is an interesting study. One of the key figures who led the removal of homosexuality from the disease classification list (DSM) by the American Psychiatric Association in 1973, and who later served as the DSM committee chair for a significant period, was Spitzer (2003), who reported on the results of conversion therapy. He reported that 64\% of male homosexuals and 43\% of female homosexuals were converted to heterosexuality through conversion therapy. Although he later expressed regret towards homosexuals who may have been hurt by his study due to persistent criticism and protests from the homosexual community, the paper was not retracted. This indicates that the submitted paper is recognized academically.

Given these scientific facts, I want to question whether it truly helps sexual minorities to deny conversion therapy without scientific evidence and remain silent about the mental and physical complications caused by homosexuality. Homosexuals who desire conversion have the right and freedom to seek help. While the American Psychological Association considers conversion therapy to be unethical, I believe it is more unethical to deny those who seek treatment the possibility of recovery. […]

\received{May 13, 2025}
\received[revised]{January 13, 2026}
\received[accepted]{April 9, 2026}

\end{document}
\endinput